\documentclass{article}

\usepackage{setspace}
\usepackage[pdftex]{graphicx}
\usepackage{caption}

\usepackage{amssymb,amsmath,amsthm}
\usepackage[margin=0.9in]{geometry} 

\newcommand{\bb}{\mathbf}

\usepackage[]{algorithm2e}

\usepackage{nicefrac}

\begin{document}
	
	\title{The Resolution Matrix for Visualizing\\ Functional Network Connectivity}
	\author{Keith Dillon}
	\maketitle


%
%
%
%
%

\begin{abstract}
	
The resolution matrix is a mathematical tool for analyzing inverse problems such as computational imaging systems.
When treating network connectivity estimation as an inverse problem, the resolution matrix describes the degree to which network nodes and edges can be resolved.
This is useful both for quantifying robustness of the network estimate, as well as identifying correlated activity.
In this report we analyze the resolution matrix for functional MRI data from the Human Connectome project.
We find that common metrics of the resolution metric can be used to identify networked activity, though with a new twist on the relationship between default mode network and the frontoparietal attention network.

%
%
%
%
%

%
%

\end{abstract}

%
%
%

%
%

\section*{Background}

Functional connectivity in the brain, i.e. functional brain networks, are increasingly believed to be important to both healthy brain function and mental illness \cite{sporns_human_2005,fornito_fundamentals_2016}. 
However, the ability to monitor brain activity in vivo is severely-limited in humans, requiring the use of indirect methods such as functional magnetic resonance imaging (fMRI).
Due to the relatively-poor resolution and signal-to-noise ratio of these functional imaging methods, the determination of network nodes and edges is a difficult and unsolved problem \cite{dillon_resolution-based_2020}.

\begin{figure}[h!] \centering 
	\scalebox{0.45}{\includegraphics[clip=true, trim=0in 0in 0in 0in]{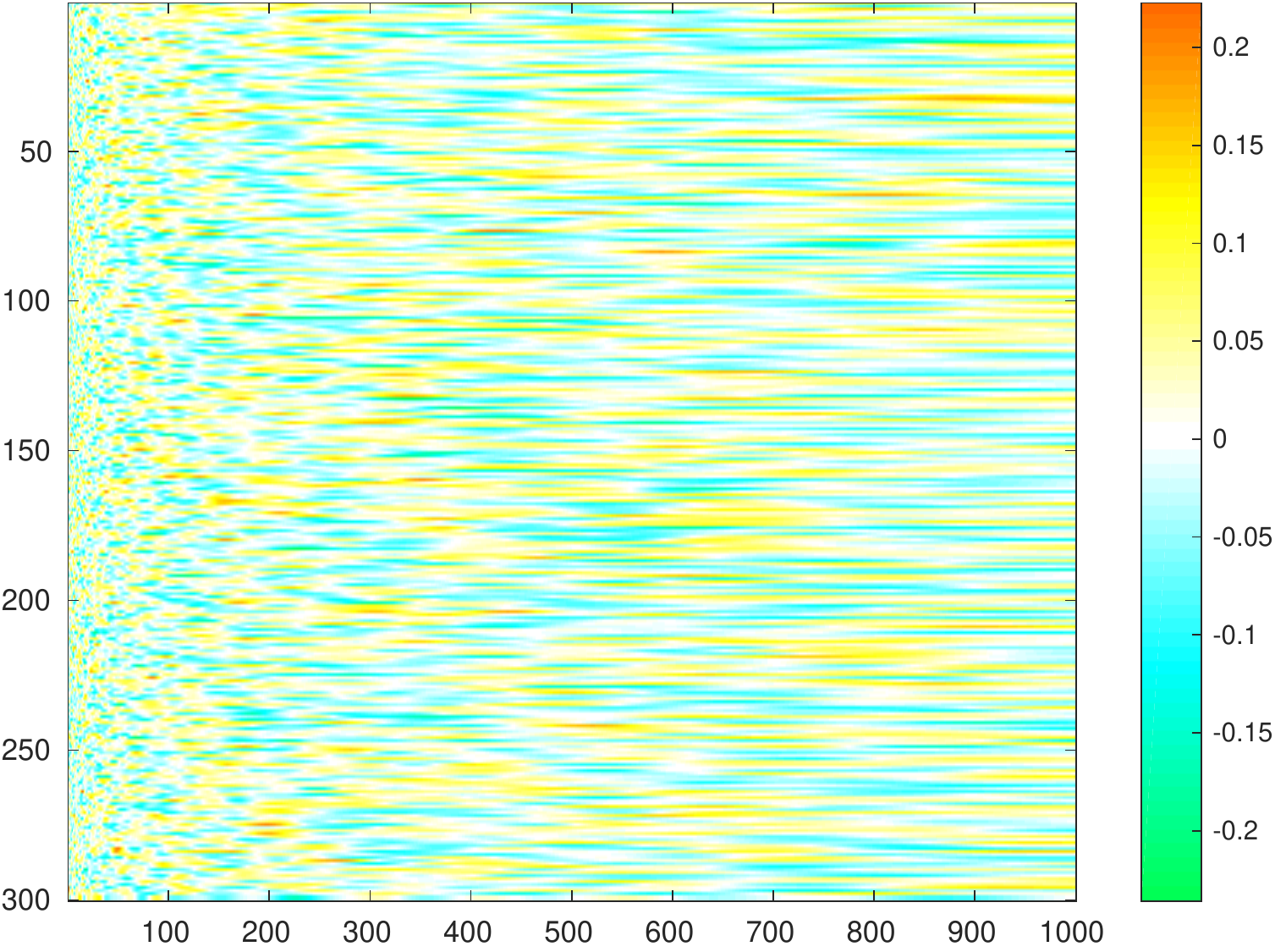}} 
	\scalebox{0.45}{\includegraphics[clip=true, trim=0in 0in 0in 0in]{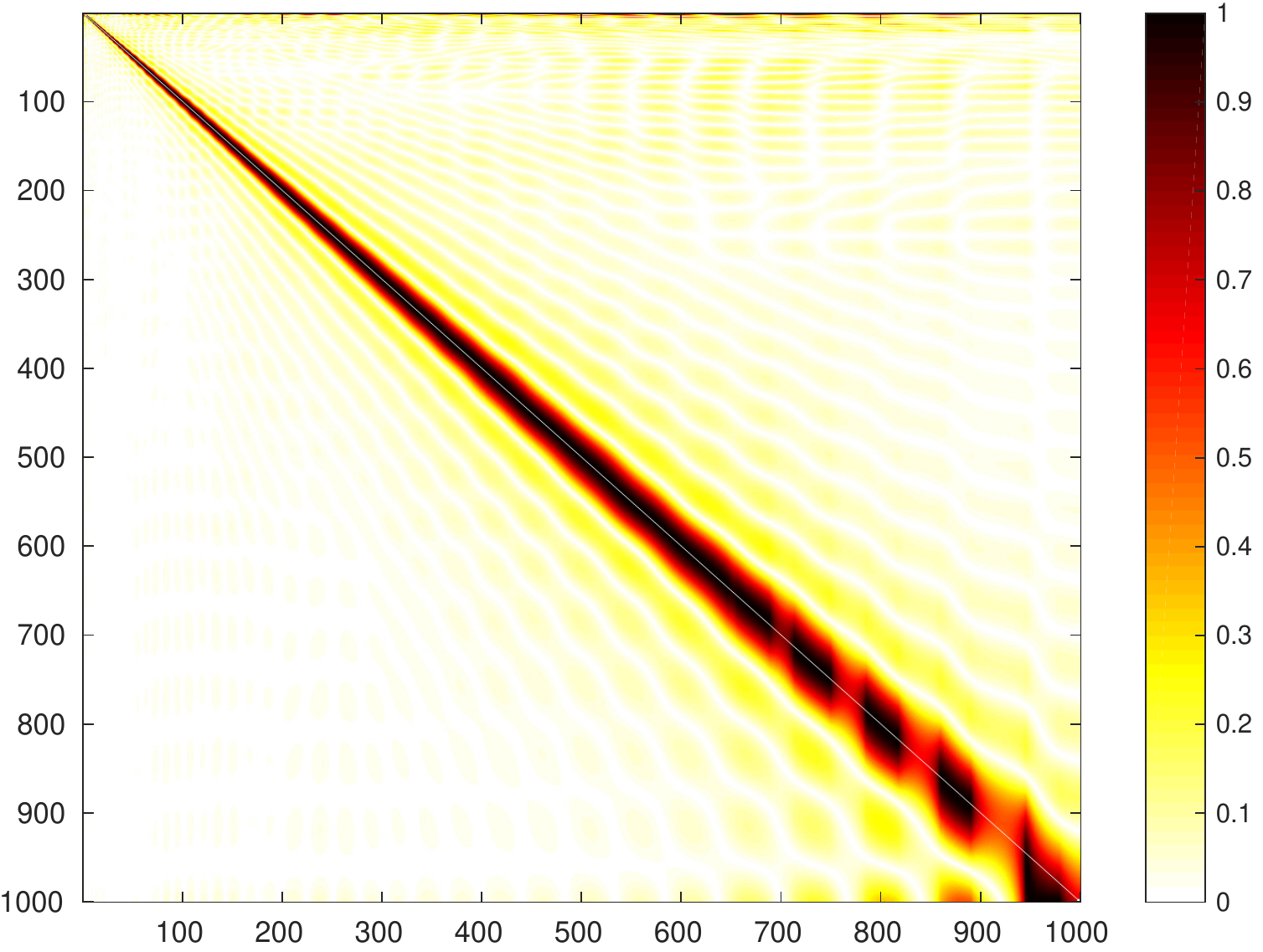}} 
	\caption{(Left) data matrix for one-dimensional simulation, where column represent the signal received from each location; the columns become increasingly blurred together from left-to-right; (Right) Corresponding resolution matrix, where each column represents the best-resolved pixel than can be formed at the corresponding location.}  
	\label{fig_resosim1}
\end{figure}

\begin{figure}[h!] \centering 
	\scalebox{0.6}{\includegraphics[clip=true, trim=0in 0in 0in 0.16in]{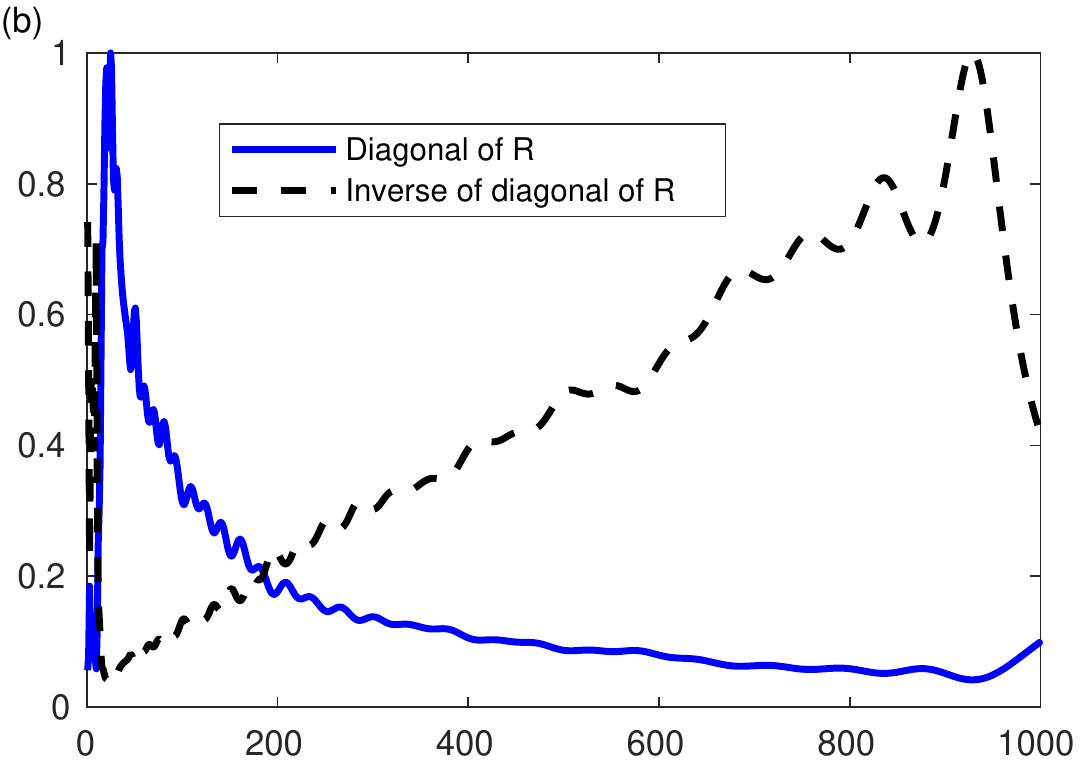}} 
	\scalebox{0.61}{\includegraphics[clip=true, trim=0in 0in 0in 0in]{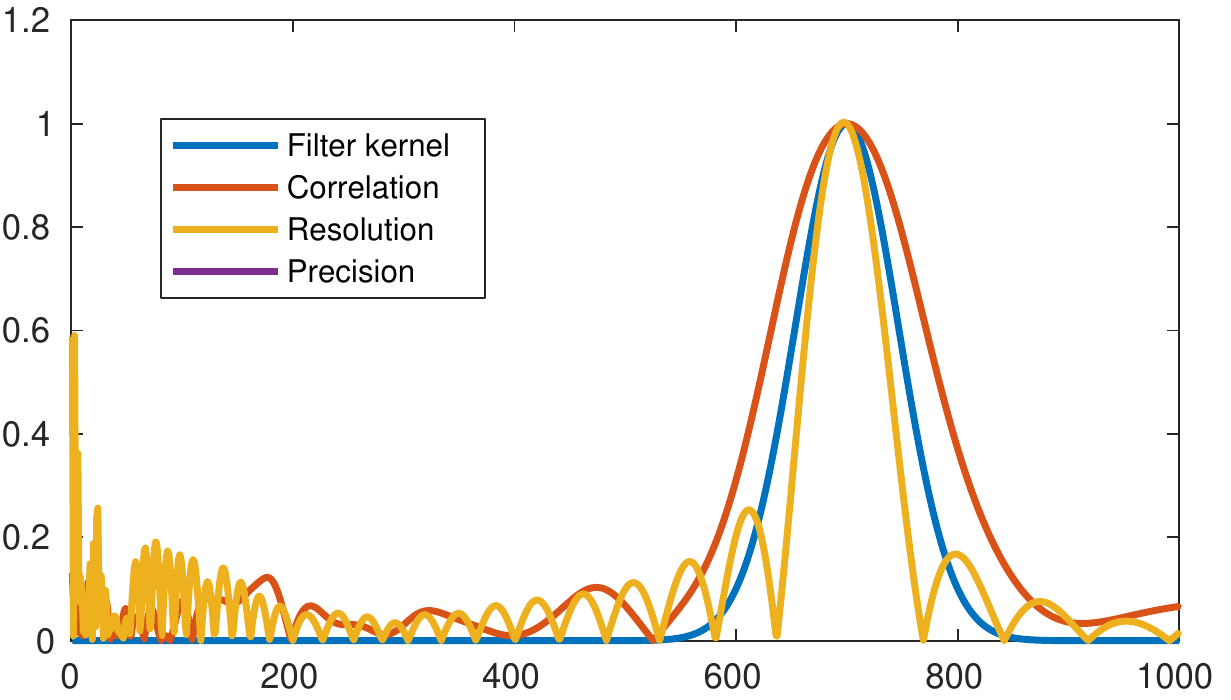}} 
	\caption{(Left) plot of the value of the resolution matrix on the diagonal versus its inverse; (Right) Plots of a single resolution cell (a resolution matrix column) compared to the original blurring kernel used at this point, the corresponding column of the correlation marix for this data matrix, and the corresponding precision matrix column (not visible because it is identical to the resolution matrix column for this case).}
	\label{fig_resosim_metrics}
\end{figure}

Resolution \cite{den_dekker_resolution:_1997} is an imaging concept that might shed some light on this problem. 
The resolution of an imaging system is the size of the smallest region who's pixel value can be determined independently of its neighbors'. 
In an inverse problem, this may be generalized to the size of the most compact region for which a unique solution may be determined \cite{dillon_element-wise_2016}.
The resolution matrix \cite{jackson_interpretation_1972} describes the smallest-resolvable regions, or resolution cells, that can be estimated at each point. 
For an ideal imaging system, the resolution cells would be individual points or small tiles even spaced across the image.
A more general case is depicted in Fig. \ref{fig_resosim1} where the resolution varies across the (one-dimensional) image and the resolution cells get correspondingly larger.
The resolution matrix does not describe the image content directly (i.e., the reflectivity or brightness parameters at points in space), but rather the size of the region over which parameter estimates are too blurred together to separate.
Fig. \ref{fig_resosim_metrics} gives the diagonal of the resolution matrix for this simulation.
The value on the diagonal is a common metric used in resolution matrix analysis because it can be efficiently computed \cite{maccarthy_efficient_2011} and relates roughly to the spread of the resolution cell (in terms of its inverse as in fig. \ref{fig_resosim_metrics}). 

Network connectivity estimation may be viewed as an inverse problem in terms of neighborhood selection, where a regression problem for each node is solved to find the connectivity to its neighbors \cite{meinshausen_high-dimensional_2006}.
In \cite{dillon_image_2016} we demonstrated the application of resolution to regression problems with functional imaging data, finding resolution cells suggestive of brain modularity or parcellation.
In \cite{dillon_resolution-based_2020} we provided a technique for individual brain parcellation by applying clustering to the resolution matrix for the neighborhood selection problem.
In \cite{dillon_computation_2019} we found a close relationship between the resolution matrix and the partial correlation estimates defining Gaussian graphical models.
This is depicted in Fig. \ref{fig_resosim_metrics}, which shows that resolvability differs from simple univariate correlation between points.
Correlation describes the similarity between signals collected from different points, which would conceptually be similar to a resolution cell describing the blurring of points together. 
However the act of computationally reconstructing the image unmixes this blurring to the degree possible, hence resolution (and partial correlation) provide a kind of ``sharpened" estimate as compared to univariate correlation. 
While the goal of resolution estimation in imaging is to identify which regions are unresolvable due to physical limitations of the imaging system, partial correlation is used to suggest network relations such as causality. 
In application to networks, therefore, resolution provides a combination of both kinds of information. 
%
%
%
%
%
In this paper we will consider the use of the resolution matrix as a visualization tool to describe these facets of a brain network.
By extracting metrics of resolution for each point, we see both how sharply a node may be defined, as well as the distant regions which are functionally similar.

\section*{Results}

We used the `10 unrelated subjects' dataset from the Human Connectome Project (HCP).
This contains multiple fMRI scans of each subject.
We used resting state scans which have been preprocessed \cite{glasser_minimal_2013} to remove artifacts, extract cortical and subcortical surfaces, and align to a standard coordinate system. 
We used the connectome workbench \cite{marcus_human_2013} to perform spatial smoothing with a 5mm kernel. 
Finally we removed extracted the signals for cortical surfaces into a data matrix $\bb A$ of size $1200\times64984$ representing 120 time samples for each of 64984 points.  

The resolution matrix for each scan is $\bb R_r = \bb A_r^\dagger \bb A$ where $\bb A_r^\dagger$ is the regularized pseudoinverse using a cutoff of 50 percent of singular values (see Appendix A).
This would produce a $64984 \times 64984$ matrix, with the $k$th column (i.e., of size $64984 \times 1$) for the $k$th point on the surface describing the resolvability of this point versus all other points on the surface. 
This is too large for conventional memory, however we can directly compute the diagonal of the resolution matrix, giving a metric for the resolution at each point as described in the previous section, by computing the sum of singular vectors squared (see Appendix A). 
In Fig. \ref{fig_avgresometric} we give the resolution metric averaged over all ten subjects for both lateral and medial views of both hemispheres.
\begin{figure}[h!] \centering 
	\scalebox{0.25}{\includegraphics[clip=true, trim=0in 0in 0in .2in]{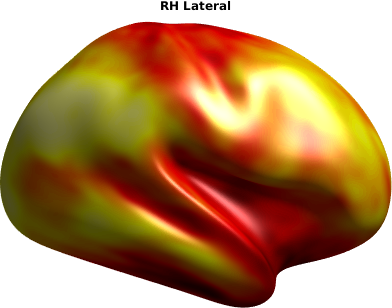}} 
	\scalebox{0.25}{\includegraphics[clip=true, trim=0in 0in 0in .2in]{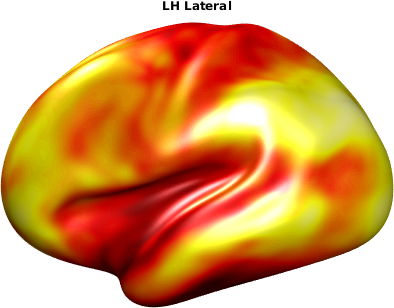}} 
	\scalebox{0.25}{\includegraphics[clip=true, trim=0in 0in 0in .2in]{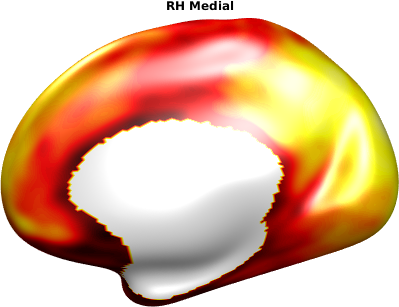}} 
	\scalebox{0.25}{\includegraphics[clip=true, trim=0in 0in 0in .2in]{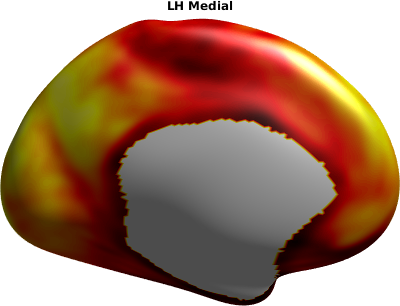}} \\ 
	\caption{Average resolution metric over 10 subjects, lateral and medial views of both cortical hemispheres. Darker agrea mean higher metric (more compact resolution cell), and lighter areas mean lower metric (more spread out resolution cell, and hence worse resolution).}
	\label{fig_avgresometric}
\end{figure}
We find that the resolution is high near the central sulcus and sylvian fissure, regions of primary cortex. 
On the medial cortex, resolution is also high on the cingulate gyrus and especially near the entorhinal cortex, where memory and emotion systems are located. 

\begin{figure}[h!] \centering 
	\scalebox{0.25}{\includegraphics[clip=true, trim=0in 0in 0in .2in]{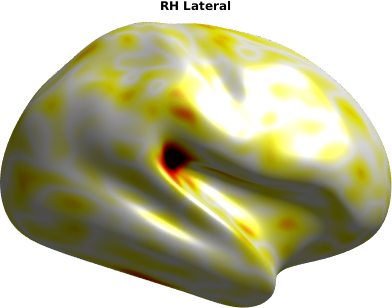}} 
	\scalebox{0.25}{\includegraphics[clip=true, trim=0in 0in 0in .2in]{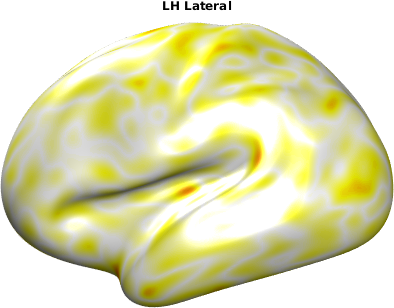}} 
	\scalebox{0.25}{\includegraphics[clip=true, trim=0in 0in 0in .2in]{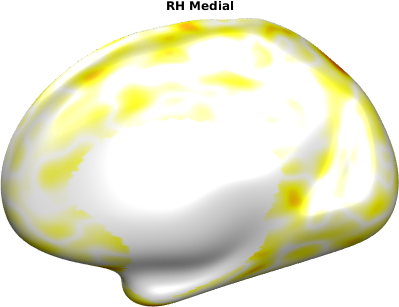}} 
	\scalebox{0.25}{\includegraphics[clip=true, trim=0in 0in 0in .2in]{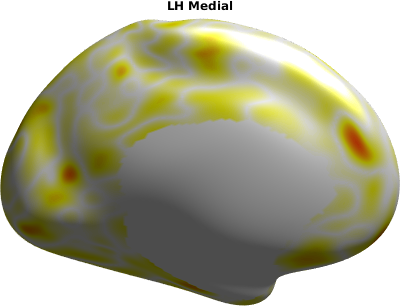}} \\
	\caption{Resolution cell for point in the right auditory cortex. A compact region around the selected point dominates the resolution cell.}		
	\label{fig_peakresocell1}	
	\scalebox{0.25}{\includegraphics[clip=true, trim=0in 0in 0in .2in]{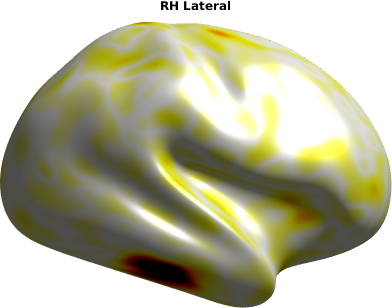}} 
	\scalebox{0.25}{\includegraphics[clip=true, trim=0in 0in 0in .2in]{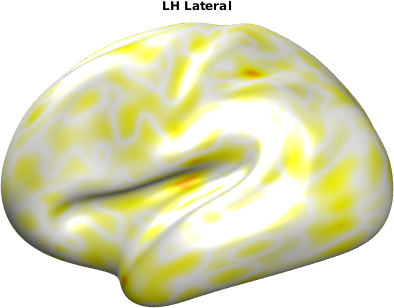}} 
	\scalebox{0.25}{\includegraphics[clip=true, trim=0in 0in 0in .2in]{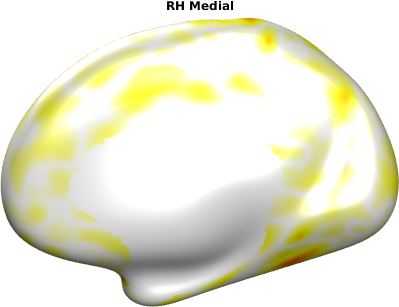}} 
	\scalebox{0.25}{\includegraphics[clip=true, trim=0in 0in 0in .2in]{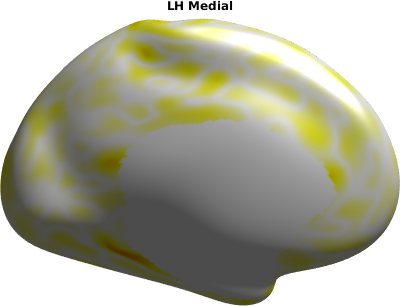}} \\
	\caption{Resolution cell for point in the right inferior temporal gyrus. A larger region around the selected point dominates the resolution cell.}		
	\label{fig_peakresocell2}			
	\scalebox{0.25}{\includegraphics[clip=true, trim=0in 0in 0in .2in]{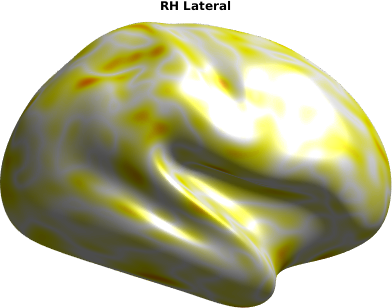}} 
	\scalebox{0.25}{\includegraphics[clip=true, trim=0in 0in 0in .2in]{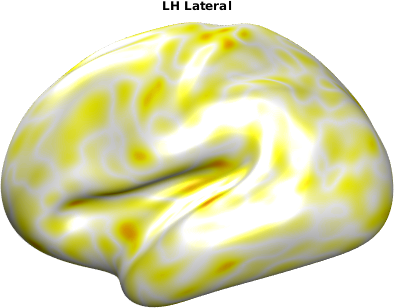}} 
	\scalebox{0.25}{\includegraphics[clip=true, trim=0in 0in 0in .2in]{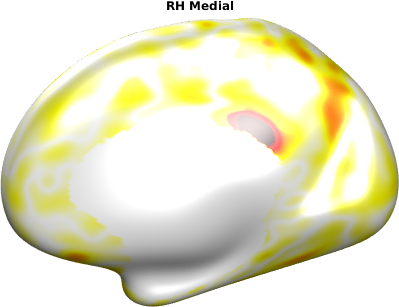}} 
	\scalebox{0.25}{\includegraphics[clip=true, trim=0in 0in 0in .2in]{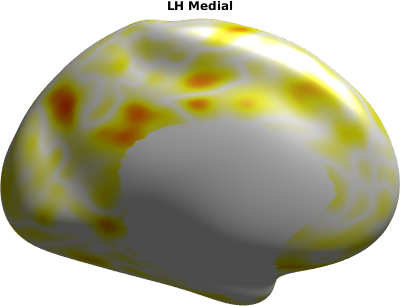}} 
	\caption{Resolution cell for point in the right congulate cortex. A compact region around the selected point dominates the resolution cell.}		
	\label{fig_peakresocell3}
\end{figure}

A high resolution metric in the above regions means the signal here is largely independent of signals in other points in the brain.
The is demonstrated in Figs. \ref{fig_peakresocell1}, \ref{fig_peakresocell2}, and \ref{fig_peakresocell3}, which show the full resolution cells for three different points on the cortex (Fig. \ref{fig_avgresometric} was an image of the peak of such resolution cells for every point).
The compact dark region around the selected point implies that nearby points are the only ones which are strongly dependent, due either to modularity of activity or resolution limits (including spatial smoothing).

As the above examples demonstrate, in the case of high resolution areas, the resolution cells are relatively simple, dominated by a large compact region. 
The resolution metric itself is potentially sufficient to identify and compare such areas. 
Such compact resolution suggests the activity at these point is highly specialized or modular, with less reliance on brain-wide networks.
On the other hand, regions of low resolution are perhaps more interesting, as they are suggestive of highly-networked processing which cannot be disentangled. 
The average inverse resolution metric over subjects is given in Fig. \ref{fig_avgresometric} to identify regions with such points.
\begin{figure}[h!] \centering 
	\scalebox{0.25}{\includegraphics[clip=true, trim=0in 0in 0in .2in]{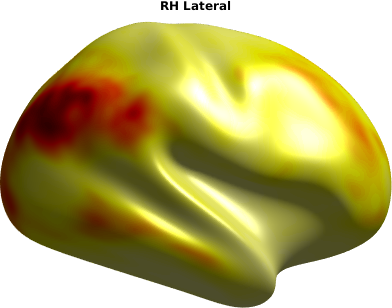}} 
	\scalebox{0.25}{\includegraphics[clip=true, trim=0in 0in 0in .2in]{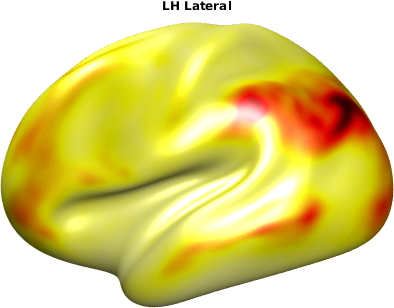}} 
	\scalebox{0.25}{\includegraphics[clip=true, trim=0in 0in 0in .2in]{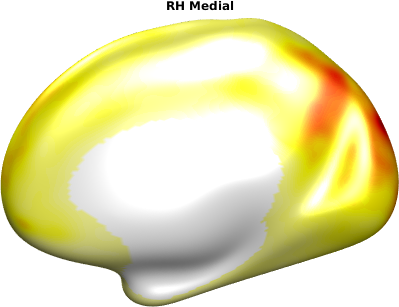}} 
	\scalebox{0.25}{\includegraphics[clip=true, trim=0in 0in 0in .2in]{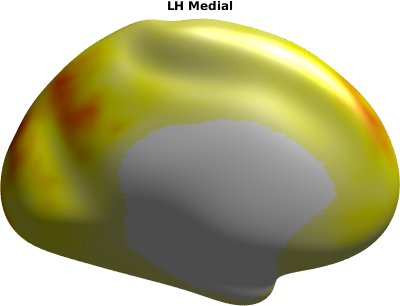}} 
	\caption{Inverse of resolution metric averaged over 10 subjects, lateral and medial views of both cortical hemispheres.}
	\label{fig_avgresometric_inv}
\end{figure}
The inverse of the resolution metric appears noticeable sparser, and dominated largely by the lateral occipital and inferioparietal region of association cortex. 
%
%

\begin{figure}[h!] \centering 
	\scalebox{0.25}{\includegraphics[clip=true, trim=0in 0in 0in .2in]{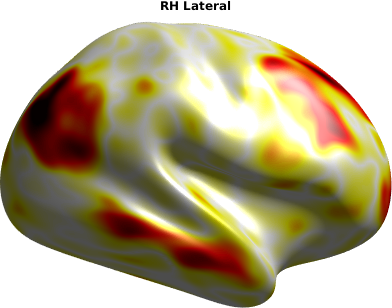}} 
	\scalebox{0.25}{\includegraphics[clip=true, trim=0in 0in 0in .2in]{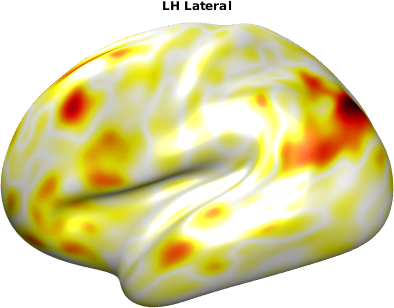}} 
	\scalebox{0.25}{\includegraphics[clip=true, trim=0in 0in 0in .2in]{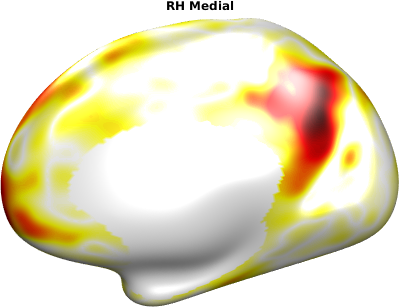}} 
	\scalebox{0.25}{\includegraphics[clip=true, trim=0in 0in 0in .2in]{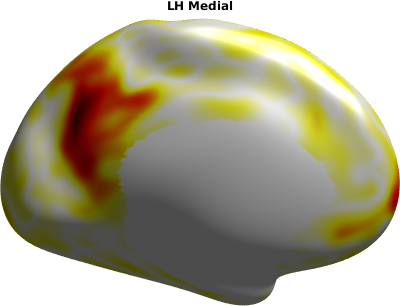}} \\
	\caption{Resolution cell for point in the right inferior parietal lobe, involving an assympetric patterns containing bilateral parietal regions, bilateral regions in the precuneus, and right superior frontal and and middle temporal gyri; exhibiting strong similarity to both frontoparietal attention network and default mode network.}	
	\label{fig_peakinvresocell1}			
	\scalebox{0.25}{\includegraphics[clip=true, trim=0in 0in 0in .2in]{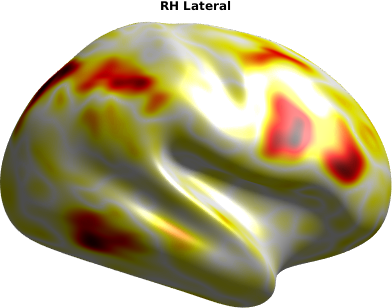}} 
	\scalebox{0.25}{\includegraphics[clip=true, trim=0in 0in 0in .2in]{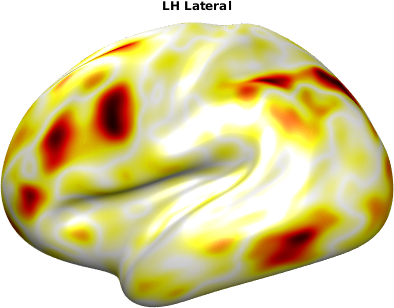}} 
	\scalebox{0.25}{\includegraphics[clip=true, trim=0in 0in 0in .2in]{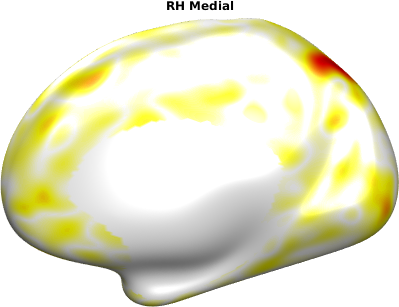}} 
	\scalebox{0.25}{\includegraphics[clip=true, trim=0in 0in 0in .2in]{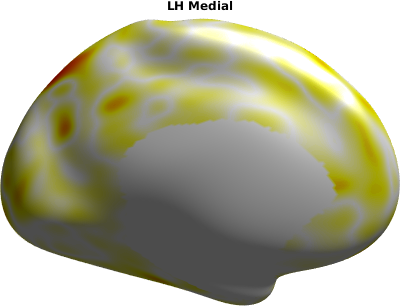}} \\
	\caption{Resolution cell for point in the right superior parietal lobe, involving a symmetric pattern of isolated lateral regions in the parietal, frontal, and temporal lobes.}	
	\label{fig_peakinvresocell2}			
	\scalebox{0.25}{\includegraphics[clip=true, trim=0in 0in 0in .2in]{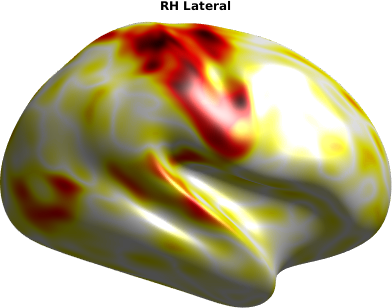}} 
	\scalebox{0.25}{\includegraphics[clip=true, trim=0in 0in 0in .2in]{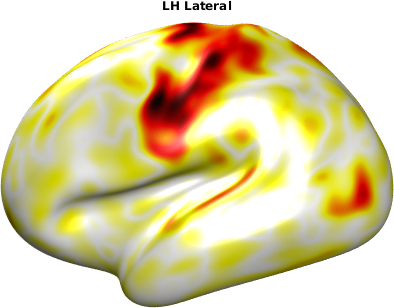}} 
	\scalebox{0.25}{\includegraphics[clip=true, trim=0in 0in 0in .2in]{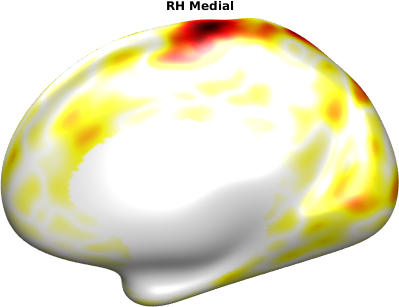}} 
	\scalebox{0.25}{\includegraphics[clip=true, trim=0in 0in 0in .2in]{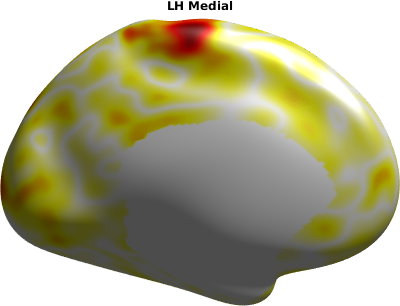}} \\
	\caption{Resolution cell for point in the post-central gyrus, involving the bilateral pre-central and post-central gyri, and a region of the occipital lobe }	
	\label{fig_peakinvresocell3}							
	\scalebox{0.25}{\includegraphics[clip=true, trim=0in 0in 0in .2in]{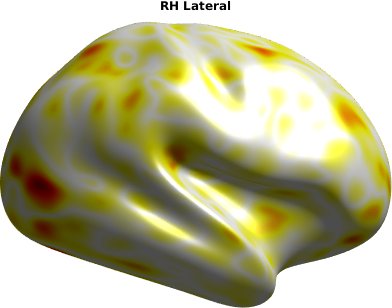}} 
	\scalebox{0.25}{\includegraphics[clip=true, trim=0in 0in 0in .2in]{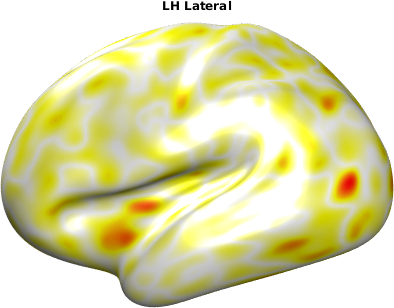}} 
	\scalebox{0.25}{\includegraphics[clip=true, trim=0in 0in 0in .2in]{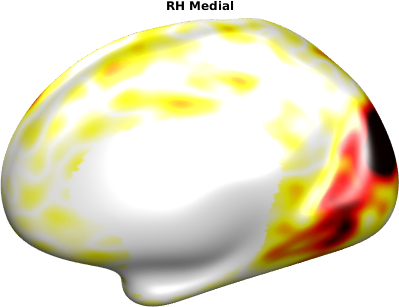}} 
	\scalebox{0.25}{\includegraphics[clip=true, trim=0in 0in 0in .2in]{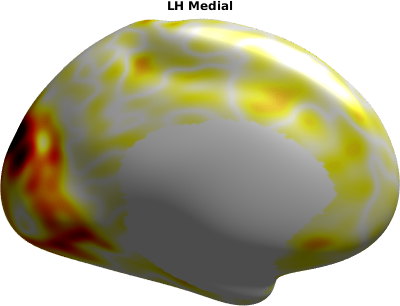}} 
	\caption{Resolution cell for point in the right cuneus, involving bilateral association regions of the medial occipital lobes.}		
	\label{fig_peakinvresocell4}				
\end{figure}

Figs. \ref{fig_peakinvresocell1} and \ref{fig_peakinvresocell2} show interesting resolution cells resulting from two points of very high inverse metric (i.e. very low resolution metric). 
Figs. \ref{fig_peakinvresocell3} and \ref{fig_peakinvresocell4} show resolution cells for points with more moderate inverse metrics, exhibiting large regional regions of sensorimotor cortex and visual association cortex, respectively.
Interestingly, the frontoparietal network \cite{marek_frontoparietal_2018} is often separated into a pair of unilateral networks \cite{cabral_exploring_2014}, while the default mode network \cite{fox_human_2005} is considered symmetric \cite{raichle_brains_2015}.
Here, we find that the resolution cell of Fig. \ref{fig_peakinvresocell1}, most similar to the default mode network, contains symmetric medial regions but asymmetric lateral regions.
Meanwhile the resolution cell of Fig. \ref{fig_peakinvresocell2} contains symmetric lateral regions but essentially no medial regions.


\section*{Discussion}

The resolution metric (and its inverse) can be used to identify points who's activity is essentially independent of other points beyond a compact region, as well as points which are involved in long-range networks of activity.
Probing these networked points more deeply, we found patterns reminiscent of well-known networks such as the default-mode and frontoparietal networks. 
These two networks in particular were originally defined as being positively-correlated, while being negatively-correlated between each other \cite{fox_human_2005}.
Further, Independent component analysis generally separated the frontoparietal network into a unilateral pair \cite{cabral_exploring_2014}.
However from the perspective of resolution cells, we find a different mix of regions and asymmetry, with a symmetric network of lateral regions, and an asymmetric network with symmetric medial regions and asymmetric lateral regions.

\clearpage\newpage

\section*{Appendix A: Resolution matrix derivations}

In the neighborhood selection problem \cite{meinshausen_high-dimensional_2006}, we solve the linear regression 
\begin{align}
\bb A \bb x_k  = \bb a_k,
\label{eq_inverse_nbhd}
\end{align}
where $\bb A$ is the matrix whos the $k$th column $\bb a_k$ contains the time series describing the fMRI activity of the $k$th cortical point. 
The predictor $\bb x_k$ can be viewed as the $k$th column of a weighted adjacency matrix $\bb X$ which relates points to each other.
The least-squares solution is 
$$\hat{\bb x}_k = \bb A^\dagger \bb a_k = \bb A^\dagger \bb A \bb x_k = \bb R \bb x_k,$$ 
where $\bb A^\dagger$ is the pseudoinverse of $\bb A$, and we have defined the resolution matrix, 
\begin{align}
\bb R = \bb A^\dagger \bb A.
\end{align}
$\bb R$ describes the difference between the least-squares solution $\hat{\bb x}_k$ and the true predictor  $\bb x_k$.
The more $\bb R$ differs from the identity matrix, the more information loss there is, and the worse our estimate of the true connectivity will be.

In regression as with other supervised machine learning techniques, there is a risk of overfitting when there is limited data and many parameters. 
In a single-subject scan from the HCP data, we have 1200 time samples for each of 64984 cortical surface points. 
An adjacency matrix relating every point to every other point would require $64984 \times 64984$ variables (one per edge), far more than the number of measurements. 
The least squares solution above is a form of regularization which presumes the missing components of the data are zero. 
However the small components are also likely to be dominated by noise due to the poor signal-to-noise ratio (SNR) of fMRI data, suggesting a need for additional regularization. 
One popular approach for dealing with this SNR is to truncate the singular value decomposition (SVD) of the data matrix \cite{dillon_resolution-based_2020}.
This can be viewed as a regularization technique for regression, which 
yields similar results to Ridge regression \cite{hansen_truncatedsvd_1987}.
In \cite{dillon_resolution-based_2020}, a cutoff of 30 to 50 percent of singular values was found by cross-validation to be optimal for regularizing the predictor for fMRI data.

The truncated SVD (TSVD) of $\bb A$ is $\bb A = \sum_{i=1}^r \sigma_i \bb u_i \bb v_i^T$, where $\bb u_i$ and $\bb v_i$ are left and right singular vectors, and $\sigma_i$ are the $r$ largest singular values. 
So the TSVD regularized solution 
is
\begin{align}
\hat{\bb x}_k = \sum_{i=1}^r \frac{1}{\sigma_i} \bb v_i \bb u_i^T \bb a_k = \bb A_r^\dagger \bb a_k,
\label{eq_tsvd_soln}
\end{align}
where we have defined the TSVD-regularized pseudoinverse $\bb A_r^\dagger$.
The resulting TSVD-regularized resolution matrix is
\begin{align}
\bb R_r = \bb A^\dagger_r \bb A_r = \bb A^\dagger_r \bb A.
\end{align}

To form an efficient estimate of the diagonal of $\bb R_r$ for use as a metric, we first note $\bb R_r$ can be written as
\begin{align}
\bb R_r = \bb A_r^\dagger \bb A 
= \left(\sum_{i=1}^r \frac{1}{\sigma_i} \bb v_i \bb u_i^T\right)
\left(\sum_{i=1}^m \sigma_i \bb u_i \bb v_i^T\right)
= \sum_{i=1}^r \bb v_i \bb v_i^T.
\label{eq_Rr_svd}
\end{align}
So the diagonal elements of $\bb R_r$ is 
\begin{align}
(\bb R_r)_{k,k} = \bb A_r^\dagger \bb A 
= \sum_{i=1}^r (\bb v_i)^2_k,
\label{eq_Rkk_svd}
\end{align}
the sum of squares of the $k$th elements of the first $r$ singular vectors (ie. corresponding to $r$ largest singular values).

We can also efficiently compute a single (i.e., the $k$th) resolution cell directly as the $k$th column of $\bb R_r$,
\begin{align}
(\bb R_r)_{k} = \bb A_r^\dagger \bb a_k 
= \sum_{i=1}^r \bb v_i (\bb v_i)_k.
\label{eq_rk_svd}
\end{align}

\section*{Appendix B: Detailed data}

\begin{figure}[h!] \centering 
	\scalebox{0.25}{\includegraphics[clip=true, trim=0in 0in 0in 0in]{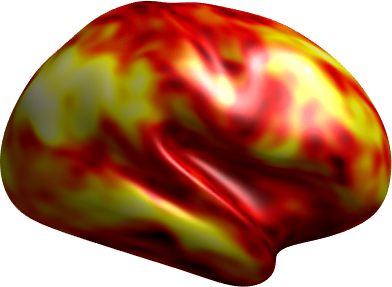}} 
	\scalebox{0.25}{\includegraphics[clip=true, trim=0in 0in 0in 0in]{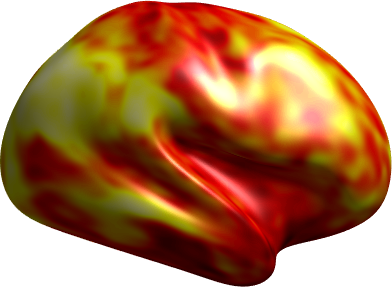}} 
	\scalebox{0.25}{\includegraphics[clip=true, trim=0in 0in 0in 0in]{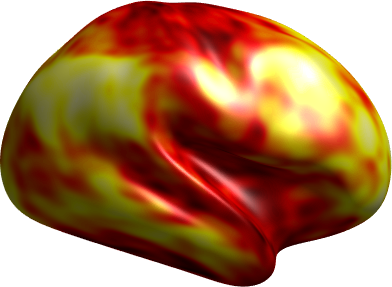}} 
	\scalebox{0.25}{\includegraphics[clip=true, trim=0in 0in 0in 0in]{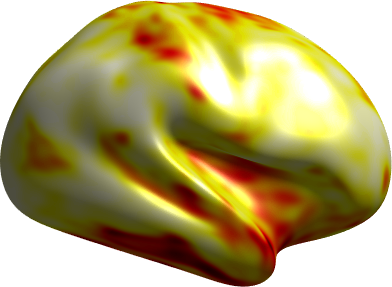}} 
	\scalebox{0.25}{\includegraphics[clip=true, trim=0in 0in 0in 0in]{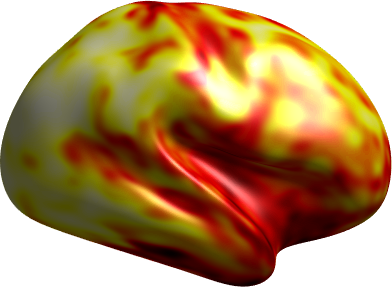}} 
	\scalebox{0.25}{\includegraphics[clip=true, trim=0in 0in 0in 0in]{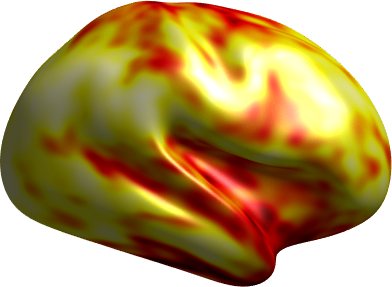}} 
	\scalebox{0.25}{\includegraphics[clip=true, trim=0in 0in 0in 0in]{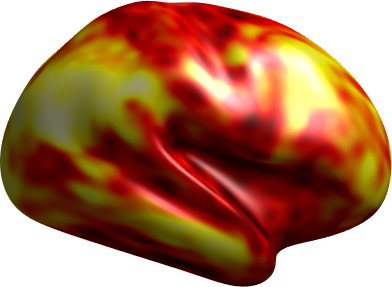}} 
	\scalebox{0.25}{\includegraphics[clip=true, trim=0in 0in 0in 0in]{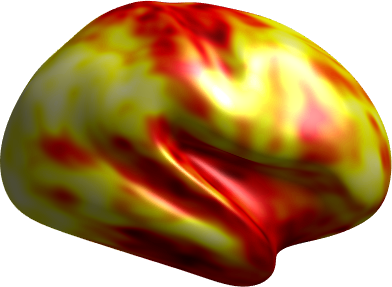}} 
	\scalebox{0.25}{\includegraphics[clip=true, trim=0in 0in 0in 0in]{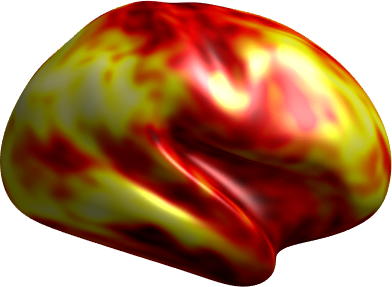}} 
	\scalebox{0.25}{\includegraphics[clip=true, trim=0in 0in 0in 0in]{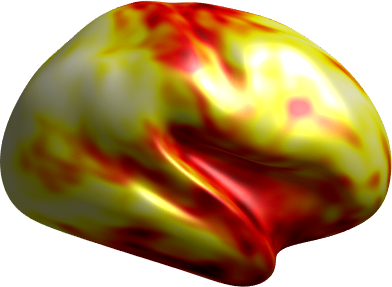}} 
	\caption{Resolution metric images for 10 subjects, right lateral view of cortex.}
	\label{fig_resometric10}
	\scalebox{0.56}{\includegraphics[clip=true, trim=0in 0in 0in .2in]{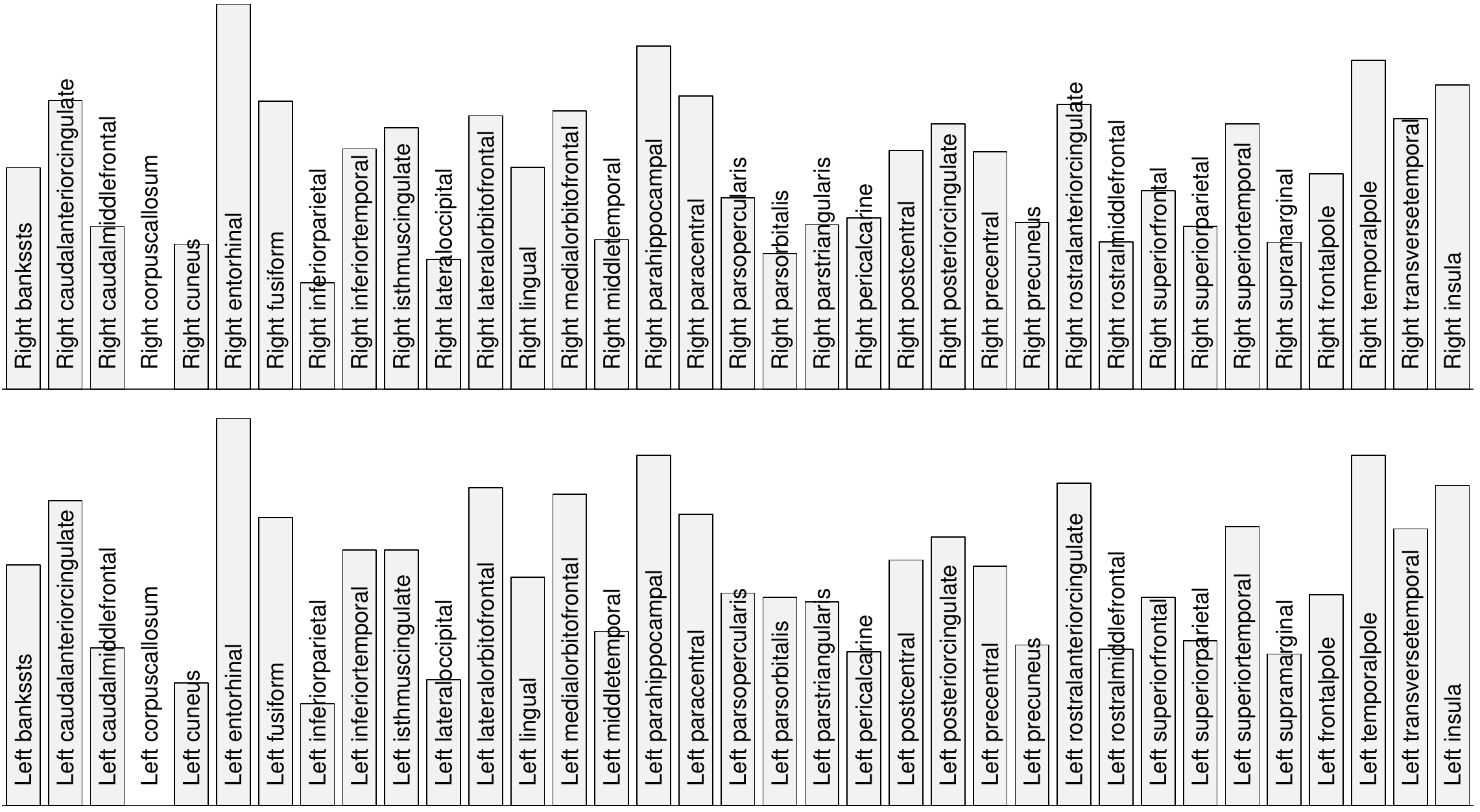}} 
	\caption{Relative contribution of regions to average metric.}
	\label{fig_resometriclabels}
	\scalebox{0.56}{\includegraphics[clip=true, trim=0in 0in 0in .2in]{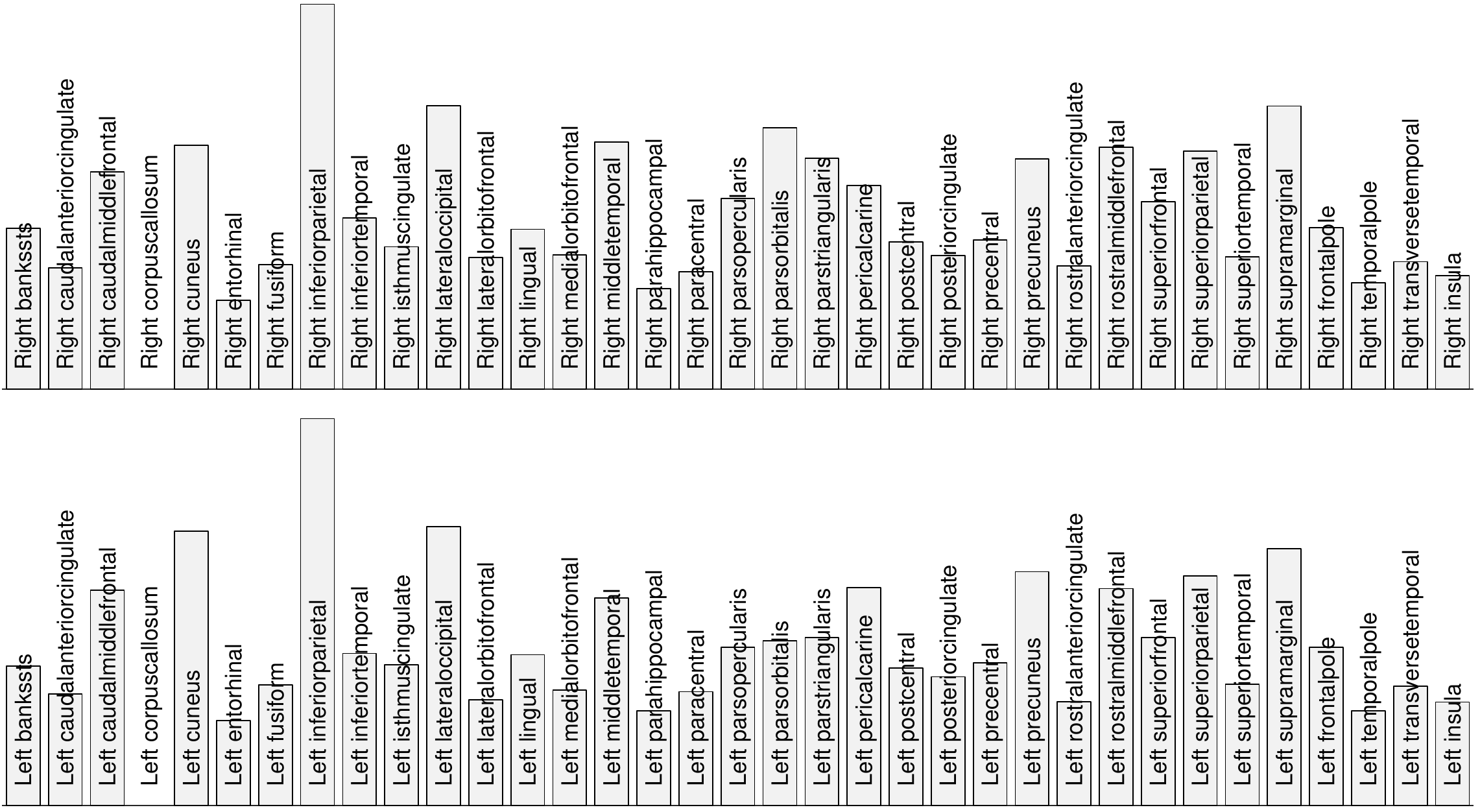}} 
	\caption{Relative contribution of regions to average inverse metric.}
	\label{fig_invresometriclabels}	
\end{figure}

Fig. \ref{fig_resometric10} provides visualizations of the resolution metric  for each subject, demonstrating that the pattern appears generally the same for each individual.
Figs. \ref{fig_resometriclabels} and \ref{fig_invresometriclabels} give the relative contribution of different regions to the resolution metric and its inverse.

%
%

\end{document}